# Different universality classes of isostructural U$TX$ compounds ($T$ = Rh, Co, Co$_{0.98}$Ru$_{0.02}$; $X$ = Ga, Al)

Petr Opletal, Vladimír Sechovský and Jan Prokleška

*Charles University, Faculty of Mathematics and Physics, Department of Condensed Matter Physics*
*Ke Karlovu 5, Prague 2, Czech Republic*

**Abstract**

Magnetization isotherms of the 5$f$-electron ferromagnets URhGa, UCoGa and UCo$_{0.98}$Ru$_{0.02}$Al were measured at temperatures in the vicinity of their Curie temperature in order to investigate the critical behavior near the ferromagnetic phase transition. These compounds adopt the layered hexagonal ZrNiAl-type structure and exhibit huge uniaxial magnetocrystalline anisotropy. The critical $\beta$, $\gamma$ and $\delta$ exponents were determined by analyzing Arrott-Noakes plots, Kouvel-Fisher plots, critical isotherms, scaling theory and Widom scaling relations. The values obtained for URhGa and UCoGa can be explained by the results of the renormalization group theory for a 2D Ising system with long-range interactions similar to URhAl reported by other investigators. On the other hand, the critical exponents determined for UCo$_{0.98}$Ru$_{0.02}$Al are characteristic of a 3D Ising ferromagnet with short-range interactions suggested in previous studies also for the itinerant 5f-electron paramagnet UCoAl situated near a ferromagnetic transition. The change from the 2D to the 3D Ising system is related to the gradual delocalization of 5$f$ electrons in the series of the URhGa, URhAl, UCoGa to UCo$_{0.98}$Ru$_{0.02}$Al and UCoAl compounds and appears close to the strongly itinerant nonmagnetic limit. This indicates possible new phenomena that may be induced by the change of dimensionality in the vicinity of the quantum critical point.

**Introduction**

Critical phenomena have been one of the most studied issues of physics since the critical points were discovered by T. Andrews [1]. The continuous (second-order) phase transition was found in systems like ferromagnets [2] to be connected with unified behavior near and at the critical point which can be described by critical exponents. In the case of ferromagnets three critical exponents, $\gamma$, $\beta$ and $\delta$, characterize the behavior near and at the critical point. They can be determined from experimental data using the relations [3]:

$$M_S(T) \sim |t|^\beta \, (T < T_C) \tag{1}$$

$$\chi(T)^{-1} \sim |t|^{-\gamma'}(T < T_C), |t|^{-\gamma}(T > T_C) \tag{2}$$

$$M_s \sim (\mu_0 H)^{1/\delta} \, for \, T = T_C, \tag{3}$$

where $M_s$ is the spontaneous magnetization, $\chi$ is magnetic susceptibility and $H$ is a magnetic field.

The universal behavior near critical points was described by the renormalization group theory first mentioned by L. P. Kadanoff [4] and then fully developed by K. G. Wilson [5–7]. The universality class is determined by dimensionality of the system $d$, dimensionality of the order parameter $n$ and the range of the interaction. There exist several universality classes in magnetism. The most known are 3D Ising ($n = 1$), XY model ($n = 2$) and 3D Heisenberg ($n = 3$).

It is desirable to investigate how a certain universal critical behavior and magnetic dimensionality is related to the particularities of a given material (symmetry of crystal structure, hierarchy and anisotropy of magnetic interactions, degree of localization of "magnetic" electrons, etc.). Large groups of isostructural compounds containing transition-element ions with one type of "magnetic" $d$ or $f$ electrons provide useful playgrounds for investigation of these aspects. The U$TX$ compounds ($T$ - transition metal, $X$ – p-metal) crystallizing in the hexagonal ZrNiAl-type structure constitute such a suitable group of materials [8]. The crystal structure consists of U-$T$ and $T$-$X$ basal plane layers alternating along the $c$-axis. The strong bonding of 5$f$-electron orbitals within the U-$T$ layer in conjunction with strong spin-orbit interaction leads to a huge uniaxial magnetocrystalline anisotropy that locks the U magnetic moments in the $c$-axis and thus makes these materials suitable for investigating Ising systems.

The critical magnetic behavior has so far been studied on two compounds of this isostructural group, UCoAl [9] and URhAl [10]. UCoAl, is an itinerant 5$f$ electron paramagnet undergoing, at low temperatures, a metamagnetic transition with a critical field of ~ 0.7 T [11]. Karube et al. reported that it behaves near to the critical endpoint as a 3D Ising system with short-range interactions [9]. On the other hand, URhAl was reported behaving as a 2D Ising ferromagnet with long-range interactions [10]. The observed difference of magnetic dimensionality of UCoAl and URhAl indicates that the common layered hexagonal crystal structure and uniaxial magnetocrystalline anisotropy shared by all compounds of the U$TX$ family ($T$ – transition metal, $X$ = Al, Ga, Sn, In) [8,11] is not a sufficient condition for sharing also a common magnetic universality class.

To test this aspect we prepared single crystals of three hexagonal U*TX* ferromagnets – UCoGa [11,13–15], URhGa [16–18] and UCo$_{0.98}$Ru$_{0.02}$Al (a close analog of UCo$_{0.99}$Ru$_{0.01}$Al [19,20]) and measured their magnetization isotherms in the neighborhood of their Curie temperature. The measured data were analyzed by investigating Arrott-Noakes plots [21,22], Kouvel-Fischer plots [23], critical isotherms, scaling theory [24] and Widom scaling relations [25] in order to determine the critical exponents and their corresponding universality classes. The first two compounds were found behaving as 2D Ising systems similar to previously reported URhAl [10] whereas the critical exponents determined for UCo$_{0.98}$Ru$_{0.02}$Al point to the 3D Ising universality class similar to UCoAl [9]. It is discussed that the change between these two classes of universality is associated with a changing degree of delocalization of 5f electrons.

**Experimental**

Single crystals of URhGa, UCoGa and UCo$_{0.98}$Ru$_{0.02}$Al were prepared from stoichiometric melts by Czochralski method using a triarc furnace. Each single crystal was wrapped in a tantalum foil and sealed in a quartz tube in evacuated to $10^{-6}$ mbar. The EDX analysis accomplished with a scanning electron microscope Tescan Mira I LMH equipped by a backscatter electrons detector confirmed the stoichiometric composition of URhGa, UCoGa and UCo$_{0.98}$Ru$_{0.02}$Al. The X-ray Laue method (Laue diffractometer of Photonic Science) shown good quality of the crystals. Samples for magnetization were cut by a wire saw to a rectangular prism shape. The *c*-axis magnetization *M* was measured using an MPMS-7-XL (Quantum Design) in fields from 0.1 T to 4 T at different temperatures in the vicinity of Curie temperature.

The values of internal magnetic field *H* were calculated as:

$$H = H_a - N \cdot M \qquad (4),$$

where *N* is the demagnetization factor. The demagnetization factor was evaluated using the formula published by Aharoni [26].

**Results and Discussion**

*URhGa and UCoGa*

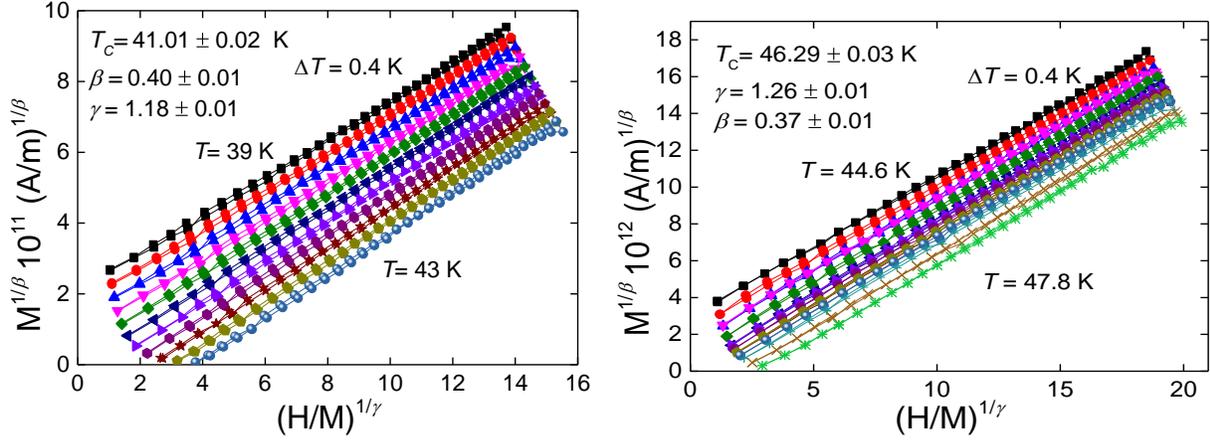

*Fig. 1: Modified Arrott plots for URhGa (left) and UCoGa (right). Resulting critical exponents and Curie temperature corresponding to each compound are in the figures. Magnetic curves close to Curie temperature were used for fit of Arrott-Noakes equation.*

The Curie temperature of a ferromagnet is usually determined by Arrott plot ($M^2$ vs $H/M$ plot) analysis of magnetization isotherms [21] measured at temperatures in the critical region. The linear Arrott plots are, in fact, a graphical representation of the Landau equation of state in the theory of second-order phase transition [27–29], which has been later specified for ferromagnets by Ginsburg [30]. This approach is certainly suitable for investigation of homogenous isotropic ferromagnets. This condition is not met for most real systems, which is documented by considerable curvature in Arrott plots. The value of Curie temperature $T_C$ can be the refined by finding the $\beta$ and $\gamma$ coefficients for which modified Arrott plots (Arrott-Noakes plots) $M^{1/\beta}$ vs. $H/M^{1/\gamma}$ are linear. The Arrott-Noakes plot comes from the analysis of magnetization curves based on the Arrott-Noakes equation [22]:

$$(H/M)^{1/\gamma} = \frac{(T-T_C)}{T_1} + (M/M_1)^{1/\beta} \quad (5),$$

where $M_1$ and $T_1$ are material constants that are temperature independent in the vicinity of the phase transition. Magnetization curves are then fitted by (5) to get the $M^{1/\beta}$ vs. $H/M^{1/\gamma}$ plots linear and parallel by varying free parameters $\beta$ and $\gamma$ while keeping the $T_1$ and $M_1$ values fixed. The results for our studied compounds are presented in Fig. 1 and Table 1 together with the best values of $T_C$ and critical exponents $\beta$, $\gamma$.

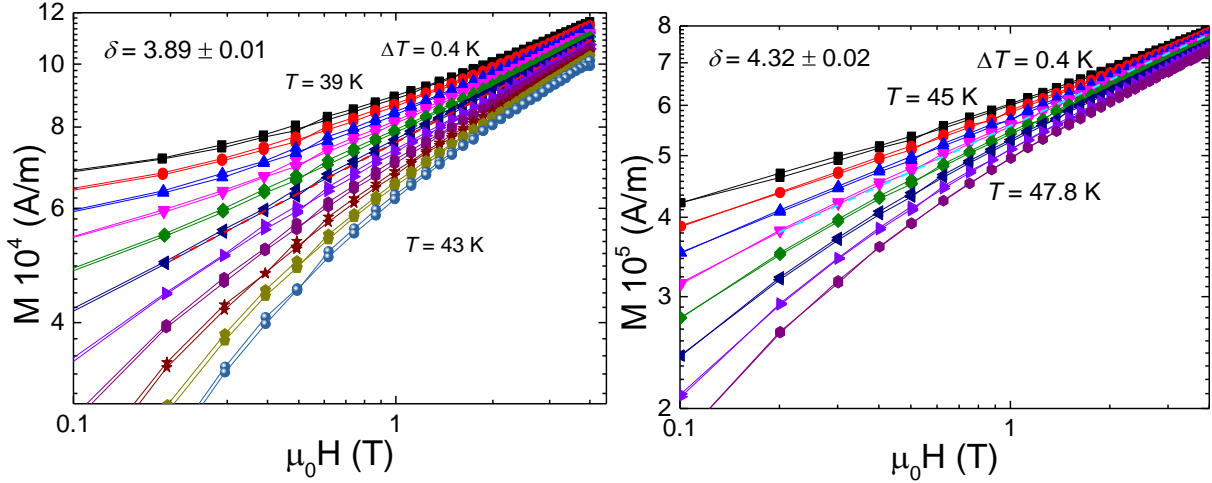

Fig. 2: Logarithmic plot of magnetic isotherms of URhGa (left) and UCoGa (right). Critical isotherms for each compound are highlighted by dashed line. The critical exponent δ from fit to critical exponent for each compound are part of the figure.

The critical exponent $\delta$ can be calculated from $\beta$ and $\gamma$ from the Widom scaling law [25]. It can be determined from the magnetization curve using Eq. (3), too. In Fig. 2 we display the magnetization curves of both compounds in a log-log plot because only the isotherm at $T_C$ is linear in this representation, as seen from Eq. (3). A linear function is then fitted to data points forming the straightest isotherm. The resulting $\delta$-values are shown in Fig. 2 and Table 1. For URhGa, UCoGa the $\delta$-value calculated from the Widom scaling law [25] is 3.95, 4.4, which is close enough to the value of 3.89 (4.5) shown in Fig. 2. The agreement between values from the Widom scaling law and values from critical isotherms could be improved as we did not know the exact $T_C$ during measurement, and therefore we used isotherms at temperature closest to the assumed $T_C$.

The $\beta$- and $\gamma$-values can be further refined by analyzing Kouvel-Fisher plots [23] which are based on the definition of critical exponents in Eqs. (1) and (2). The spontaneous magnetization $M_S$ is determined from the intersection of the $M^{1/\beta}$ axis of a Arrott-Noakes plot with straight lines and the inverse susceptibility $\chi^{-1}$ is determined from the intersections of straight lines with the $(H/M)^{1/\gamma}$ axis of the Arrott-Noakes plot. Kouvel and Fisher [23] showed that by dividing by temperature derivatives of Eqs. (3) and (4) one gets a new set of equations

$$M_S(T)[dM_S(T)/dT]^{-1} = |T - T_C|/\beta(T) \qquad (6)$$

$$\chi^{-1}(T)[d\chi^{-1}(T)/dT]^{-1} = |T - T_C|/\gamma(T). \qquad (7)$$

In the vicinity of $T_C$ the $\beta(T)$ and $\gamma(T)$ became equal to the corresponding critical $\beta$ and $\gamma$ values. The critical exponents are determined from the slope and the Curie temperature is determined from the intersection with the $T$ axis in the Kouvel-Fisher plots shown in Fig. 3. The resulting critical exponents for URhGa and UCoGa are presented in Fig. 3 and Table 1.

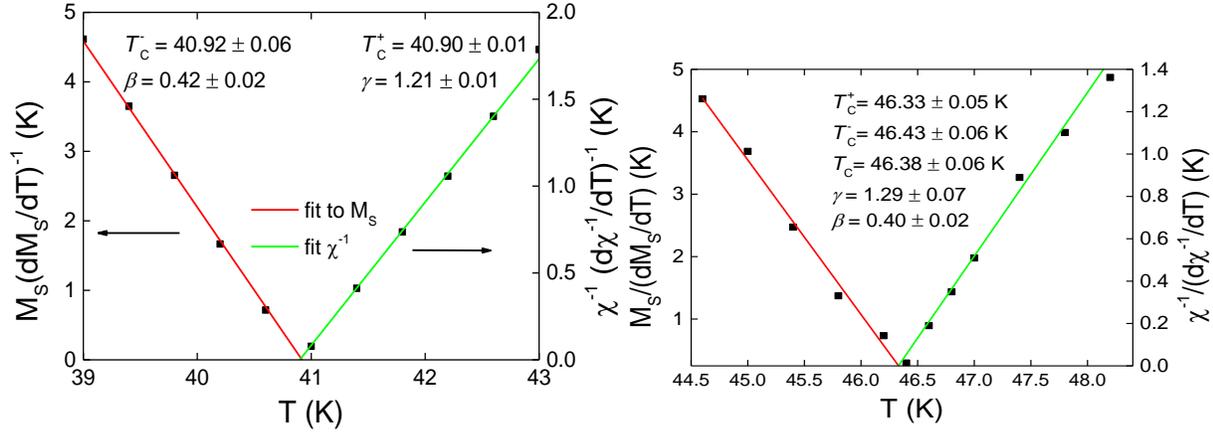

*Fig. 3: Kouvel-Fisher plot for spontaneous magnetization and susceptibility for URhGa (left) and UCoGa (right). The critical exponents are shown in corresponding figures.*

To check whether the critical exponents above and below the phase transition are equal, we have separately determined the critical exponent $\gamma$ ($T > T_C$) and $\gamma'$ ($T < T_C$) using the scaling theory that predicts a reduced equation of state close to the phase transition [24]:

$$M(\mu_0 H, t) = |t|^\beta f_\pm(\mu_0 H/|t|^{\beta+\gamma}), \qquad (8)$$

where $f_+$ is for $T > T_C$ and $f_-$ is for $T < T_C$ and both are regular analytical functions. If the correct $\beta$, $\gamma$ and $T_C$ values are chosen then the data points in the plot $M(\mu_0 H,t)/|t|^\beta$ versus $\mu_0 H/|t|^{\beta+\gamma}$ should fall on two universal curves, one for $T < T_C$ and second for $T > T_C$ and these curves should approach each other asymptotically. Magnetization data for both compounds were fitted by Eq. (8). The results are shown in Fig. 4 together with the values of critical exponents and $T_C$, respectively. For both compounds, the difference of the determined critical exponent $\gamma$ for $T$ above and below $T_C$ is very small which corroborates the presumption that it does not change.

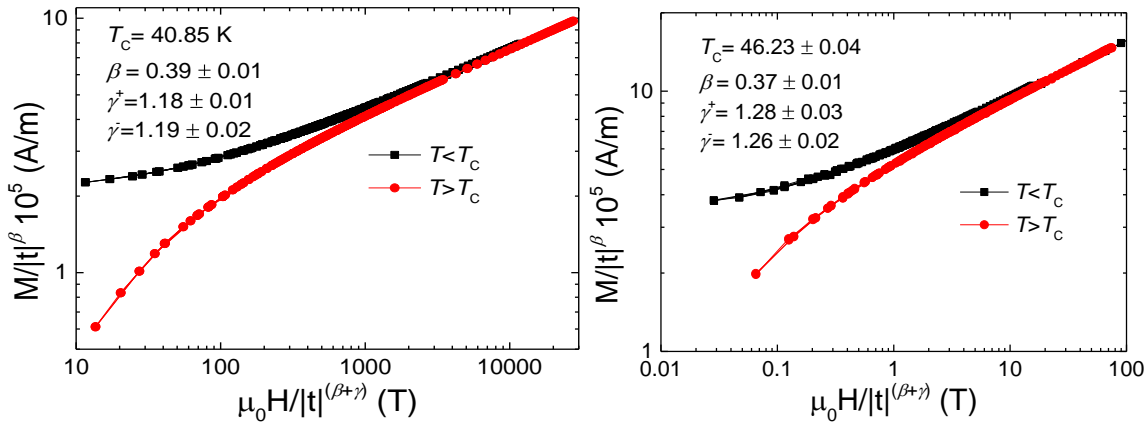

*Fig. 4: The scaled magnetization plotted against the renormalized magnetic field below and above $T_C$. Results of fit of scaling theory from Eq. (8) for both compounds are shown in figures URhGa (left) and UCoGa (right).*

The critical exponents of URhGa and UCoGa can be explained similarly to the URhAl case [10] by introducing a weak long-range magnetic exchange interaction in form $J(r) \sim r^{-(d+\sigma)}$, where $\sigma$ is the range of the exchange interaction [31]. Fischer et al. [31] applied the renormalization group theory for a system with an interaction $J(r)$ for which the equation for $\gamma$ has been obtained in the form:

$$\gamma = 1 + \frac{4}{d}\left(\frac{n+2}{n+8}\right)\Delta\sigma + \frac{8(n+2)(n-4)}{d^2(n+8)^2}\left[1 + \frac{2G\left(\frac{d}{2}\right)(7n+20)}{(n-4)(n+8)}\right]\Delta\sigma^2, \qquad (9)$$

where $\Delta\sigma = \left(\sigma - \frac{d}{2}\right)$ and $G\left(\frac{d}{2}\right) = 3 - \frac{1}{4}\left(\frac{d}{2}\right)^2$. The obtained values of $\gamma$ were examined using Eq. (9) by substituting the possible values of $d$ (dimension of the system) = 1, 2, or 3, and $n$ (dimension of the order parameter) = 1, 2, or 3 in all possible combinations and comparing the resulting values of $\sigma$ for $\gamma$ and $\beta$. The best agreement for both, URhGa and UCoGa, has been found for a 2D Ising system with LR interactions. The resulting $\sigma$ value and corresponding $\beta$, $\gamma$ and $\delta$ values are listed in Table 1. The same universality class was reported for URhAl [10].

*UCo$_{0.98}$Ru$_{0.02}$Al*

The critical exponents $\gamma = (1.27 \pm 0.01)$ and $\beta = (0.32 \pm 0.01)$ determined by Arrott-Noakes analysis of magnetization data collected on the UCo$_{0.98}$Ru$_{0.02}$Al single crystal (see Arrott-Noakes plots in Fig. 5) with $T_C = 22.79$ K compare well to the theoretical values $\gamma = 1.241$ and $\beta = 0.325$ for the 3D Ising model [24,32].

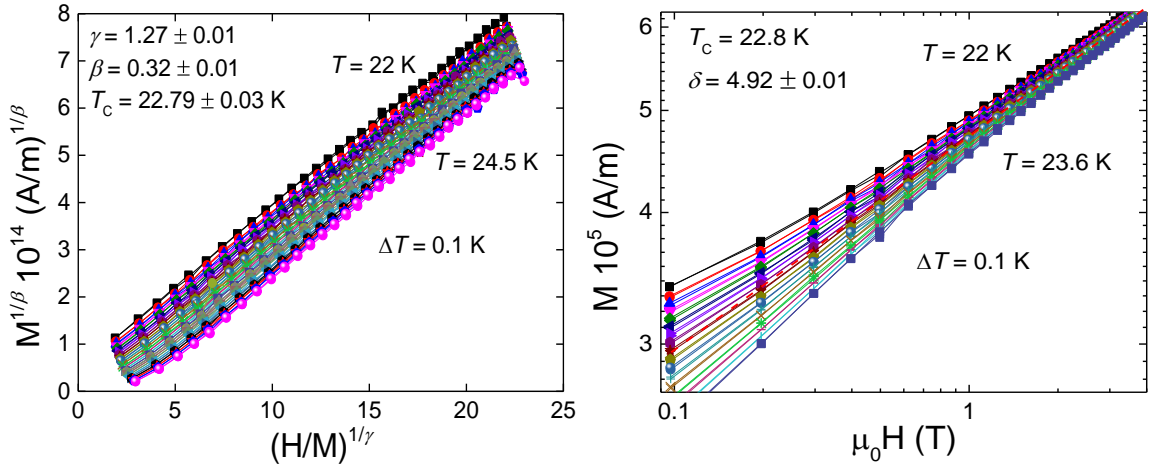

Fig. 5: Arrott-Noakes plots (left) and logarithmic plots (right) of magnetization isotherms for UCo$_{0.98}$Ru$_{0.02}$Al. The results of the fit by Arrott-Noakes equation (Eq. (5)) are displayed in left figure while in right figure the result of the fit by critical isotherm can be found.

Further on, the value $\delta = (4.92 \pm 0.01)$ corresponding to the critical isotherm (see Fig. 5) is close to the value of 4.97 determined from Widom scaling relations using $\gamma$ and $\beta$ determined from Arrott-Noakes equation. The value corresponding to the critical isotherm is close to the value of 4.82 known for the 3D Ising model.

To further analyze the UCo$_{0.98}$Ru$_{0.02}$Al magnetization data, the Kouvel-Fisher method (for the respective plot see Fig. 6) was used in a similar way as above. The resulting critical exponents $\gamma = (1.24 \pm 0.04)$ and $\beta = (0.33 \pm 0.01)$ with $T_C = (22.79 \pm 0.02)$ compare to values for the 3D Ising model.

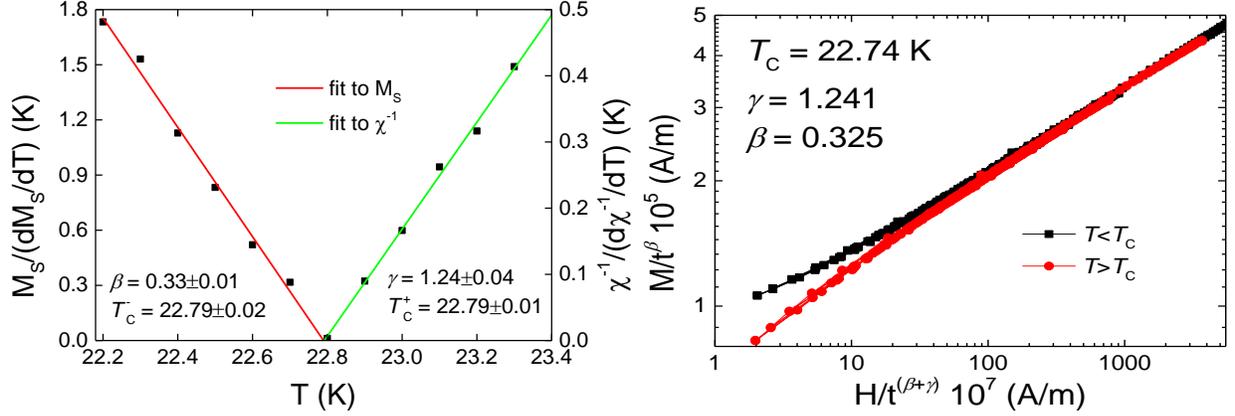

*Fig. 6: (Left) Kouvel-Fisher plot of UCo$_{0.98}$Ru$_{0.02}$Al with results of fit of Eqs. (6) and (7). (Right) Plot of the scaled magnetization vs. the rescaled magnetic field for uCo0.98Ru0.02Al. Results of fit of scaling theory from Eq. (8) are part of the figure.*

We have confirmed that UCo$_{0.98}$Ru$_{0.02}$Al behaves as a 3D Ising system by plotting magnetization data using the reduced equation of state in eq. 8 with values of $\beta$ and $\gamma$ for the 3D Ising model in Fig. 6. The data points fall on two different curves, for $T$ below and above $T_C$ = 22.74 K, respectively, which approach asymptotically to merge. The 3D Ising behavior UCo$_{0.98}$Ru$_{0.02}$Al resembles behavior of pure UCoAl reported by Karube et al. in [9].

The values of critical exponents and other relevant information obtained in the present work on URhGa, UCoGa and UCo$_{0.98}$Ru$_{0.02}$Al single crystals are displayed in Table 1. The information available for UCoAl [9] is included for comparison and further discussion.

*Table I: Critical exponents and other relevant parameters obtained for URhGa, UCoGa and UCo$_{0.98}$Ru$_{0.02}$Al by the analysis of Arrott-Noakes plots (A-N plots), Kouvel-Fisher plots (K-F plots), scaling relations (Scaling), critical isotherms (Critical i.) and data for UCoAl [9] for comparison.*

|  | Method | $T_C$ (K) | $\beta$ | $\gamma'(T<T_C)$  $\gamma(T>T_C)$ | $\delta$ | $\sigma$ |
|---|---|---|---|---|---|---|
| URhGa | A-N plots | 41.01 ± 0.02 | 0.40 ± 0.01 | 1.18 ± 0.01 | | |
|  | K-F plots | 40.91 ± 0.04 | 0.42 ± 0.02 | 1.21 ± 0.01 | | |
|  | Scaling | 40.85 ± 0.02 | 0.39 ± 0.01 | 1.18 ± 0.01  1.19 ± 0.02 | | |
|  | Critical i. | | | | 3.89 ± 0.01 | |
| LR exchange: $J(r) \sim r^{-(d+\sigma)}$ $d=2, n=1$ | | | 0.39 | 1.18 | 4.03 | 1.21 |
| UCoGa | A-N plots | 46.29 ± 0.03 | 0.37 ± 0.01 | 1.26 ± 0.01 | | |
|  | K-F plot | 46.38 ± 0.06 | 0.40 ± 0.02 | 1.29 ± 0.07 | | |
|  | Scaling | 46.23 ± 0.04 | 0.37 ± 0.01 | 1.28 ± 0.03  1.26 ± 0.02 | | |
|  | Critical i. | | | | 4.32 ± 0.01 | |
| LR exchange: $J(r) \sim r^{-(d+\sigma)}$ $d=2, n=1$ | | | 0.36 | 1.26 | 4.5 | 1.28 |

| | | | | | |
|---|---|---|---|---|---|
| UCo$_{0.98}$Ru$_{0.02}$Al | A-N plots | 22.79 ± 0.03 | 0.32 ± 0.01 | 1.27 ± 0.01 | |
| | K-F plots | 22.79 ± 0.02 | 0.33 ± 0.01 | 1.24 ± 0.04 | |
| | Scaling | 22.74 ± 0.02 | 0.325 | 1.241 | |
| | Critical i. | | | | 4.92 ± 0.01 |
| UCoAl | NMR meas. | - | 0.26 | 1.2 | 5.4 |

The table presents the most striking result of our study - although the three studied U*TX* ferromagnets adopt the same type of layered hexagonal crystal structure with strong uniaxial magnetocrystalline anisotropy they do not fall in the same universality class. URhGa and UCoGa exhibit a 2D Ising character similar to URhAl [10] whereas UCo$_{0.98}$Ru$_{0.02}$Al behaves as ta 3D Ising system, also reported for UCoAl [9].

The 5f-electrons in U intermetallics are known to have a dual character (partially localized, partially itinerant) [33–35]. The localized and itinerant characters appear in different proportions depending on crystallographic and chemical environments of U ion being reflected in a wide range of their magnetic behavior. This is a consequence of the wide extension of the 5f-wave functions which allows for considerable direct overlaps between 5f-wave functions of the nearest-neighbor U ions as well as hybridization of valence electron states of ligands (5f-ligand hybridization). As a result, the original atomic character of the 5f-wave functions is destroyed while the related magnetic moments is washed out and adequately reduced. In the strong 5f-5f overlap and 5f-ligand hybridization limits the 5f-electrons are predominantly itinerant, the 5f magnetic moments vanish and the magnetic order is lost (UCoAl in our case).

Rhodes and Wohlfarth proposed that the ratio $\mu_{eff}/\mu_s$ between the effective and spontaneous magnetic moment can be taken as a measure of the degree of itinerancy of the magnetic electrons [36,37]. In Table 2, we can see that $\mu_{eff}/\mu_s$ increases along the series as listed from top to bottom. In the Rhodes-Wohlfarth scenario, the degree of itinerancy of 5f-electrons increases when proceeding from URhGa towards UCoAl.

We can also see that the lattice parameter *a* shown in the same table simultaneously decreases along the series. The close packing of U and *T* atoms in the basal plane of the hexagonal ZrNiAl-type structure results both in a non-negligible 5f-5f overlap and a strong 5f-d hybridization involving the transition metal d states which compress most of the 5f charge density towards the basal plane. The reduction of *a* is intimately connected with the decreasing U-U and U-*T* interatomic distances within the basal plane. The simultaneously enhanced 5f-5f overlaps and 5f-d hybridization causes a higher degree of itinerancy, which corroborates the Rhodes-Wohlfarth scenario.

The change from 2D to three dimensional behavior was described by Takahashi, who considers quasi- itinerant ferromagnets in [38]. In a quasi-2D system the spin fluctuation in the *z*-direction differs from spin fluctuations in the *xy* plane, which comes from the uniaxial magnetocrystalline anisotropy of the system. The critical behavior of the system in the region of the *T-H* space centered on [$T_C$, 0] is described as 3D, while outside of this region the critical behavior is 2D. For stronger anisotropy, the region of 3D behavior is reduced. The large magnetocrystalline anisotropy observed in U*TX* compounds crystallizing in the ZrNiAl-type structure [8,11,15,39] leads to suppression of a 3D behavior and instead a 2D behavior is observed as seen in URhGa, UCoGa and URhAl. It is important to emphasize that the systems described by Takahashi show different critical behaviors, but the effect of anisotropy on the nature of fluctuations near $T_C$ is expected to be similar.

*Table 2: The values of effective and spontaneous magnetic moment, $\mu_{eff}$ and $\mu_s$, respectively, and the $\mu_{eff}/\mu_s$ ratio and lattice parameters for our studied URhGa, UCoGa and UCo$_{0.98}$Ru$_{0.02}$Al compounds completed by the values for URhAl and UCoAl. The "$\mu_s$ value" for UCoAl is the magnetic moment in the field just above the metamagnetic transition. The true $\mu_s$ value for UCoAl is indeed equal to 0; the ground state is paramagnetic.*

| Compound | $\mu_{eff}$ ($\mu_B$/f.u.) | $\mu_s$ ($\mu_B$/f.u.) | $\mu_{eff}/\mu_s$ | Ref. | Universality class | $a$ (pm) | $c$ (pm) | Ref. |
|---|---|---|---|---|---|---|---|---|
| URhGa | 2.45 | 1.17 | 2.11 | [40] | 2D Ising | 700.6 | 394.5 | [41] |
| URhAl | 2.50 | 1.05 | 2.38 | [10] | 2D Ising | 696.5 | 401.9 | [41] |
| UCoGa | 2.40 | 0.65 | 3.69 | [15] | 2D Ising | 669.3 | 393.3 | [41] |
| UCo$_{0.98}$Ru$_{0.02}$Al | 1.73 | 0.36 | 4.81 | * | 3D Ising | 669.1 | 396.6 | * |
| UCoAl | 1.60 | 0.30 | 5.33 | [11] | 3D Ising | 668.6 | 396.6 | [42] |

*) unpublished data

Further inspection of Table 2 reveals that the change from 2D Ising to 3D Ising universality class happens when the 5f-electrons become considerably itinerant. This agrees with Takahashi's description in [38], in which for more itinerant systems larger anisotropy is needed to suppress the region of 3D behavior compared to more localized systems. This explains the 3D behavior of UCoAl and UCo$_{0.98}$Ru$_{0.02}$Al even though the magnetocrystalline anisotropy is comparable to that in URhGa, UCoGa and URhAl [8,11,15,39]. Also, we see that the p-metal affects the degree of localization/itineracy, which in turn affects the universality class of the system. The hierarchy of exchange interactions will undoubtedly play an essential role in controlling dimensionality. The involvement of theorists in resolving these issues is strongly desirable. It would also be interesting to investigate the evolution of critical exponents and magnetic dimensionality together with the development of lattice parameters while applying hydrostatic pressure. A possible change of magnetic dimensionality with applied pressure near to a quantum critical point may open new questions in the quantum criticality research.

**Conclusions**
The magnetization isotherms of their URhGa, UCoGa and UCo$_{0.98}$Ru$_{0.02}$Al were measured at temperatures near to their ferromagnetic transitions where we investigated the critical behavior. The values of critical exponents $\beta$, $\gamma$ and $\delta$ have been determined by analyzing magnetization data presented in Arrott-Noakes plots, Kouvel-Fisher plots, critical isotherms, scaling theory and Widom scaling relations. The results point to the 2D Ising universality class for URhGa and UCoGa, similar to URhAl reported by other investigators [10]. Data obtained for UCo$_{0.98}$Ru$_{0.02}$Al are characteristic of the 3D Ising universality class which was suggested in previous studies also for the itinerant 5f-electron paramagnet UCoAl. At the high degree of itineracy of 5f-electrons a change from the 2D to the 3D character is observed between UCoGa and UCo$_{0.98}$Ru$_{0.02}$Al. Possible new phenomena may be expected when a dimensionality change happens in the vicinity of the quantum critical point.

**Acknowledgments**

This research was supported by Grant Agency of Charles University, grant No. 1630218 and by the Czech Science Foundation, grant No.16-06422S. Experiments were performed in MGML (www.mgml.eu), which is supported within the program of Czech Research Infrastructures (project no. LM2018096). It was also supported by OP VVV project MATFUN under Grant CZ.02.1.01/0.0/0.0/15_003/0000487. We would also like to thank to Dr. Ross Colman for proofreading of the text, and language corrections.